# 3D Microstructure Characterization of Cu-25Cr Solid State Sintered Alloy using X-ray Computed Tomography and Machine Learning Assisted Segmentation

Lucas Varoto [1,2*], Jean-Jacques Blandin[2], Pierre Lhuissier[2], Sophie Roure[1], Anthony Papillon [1], Mélissa Chosson[1], Guilhem Martin [2]

[1] Schneider Electric Industries, F-38000 Grenoble, France
[2] Univ. Grenoble Alpes, CNRS, Grenoble INP, SIMaP, F-38000 Grenoble, France

e-mail: lucas.varoto@grenoble-inp.fr

**Abstract**
Cu-Cr-based alloys with Cr content from 5 to 50 wt.% are widely used as electrical contacts for vacuum interrupters for medium voltage applications because of their excellent combination of mechanical, thermal, and electrical conductivity. Cu-Cr electrical contacts are usually processed by sintering or casting processes. For solid-state sintered Cu-Cr materials, the physical properties vary as a function of the Cr content, phase morphology and porosity volume fraction. Some studies have investigated the effect of the microstructural characteristics of Cu-Cr alloys with different Cr content and morphology on their properties. However, the porosity characterization and Cr spatial distribution and how they affect these alloys' physical properties are not as well documented. In this study, we report an in-depth 3D characterization of the porosity and Cr-phase of solid-state sintered Cu-25Cr alloys with three final relative densities using X-ray Computed Tomography (XCT). An image analysis algorithm assisted by a machine learning-based segmentation method has been specifically developed. Results show that for Cu-25Cr solid sintered alloys there are mainly two types of pores, pores located at the Cu/Cr interfaces, and pores within the Cu matrix. The interfacial porosity represents the larger volume fraction, over 75% of the total porosity for all cases, forming a large network of interconnected pores. With the increase of final density, the Cu-matrix becomes nearly fully dense while interfacial pores still represent the largest fraction decreases in size and volume. These interfacial pores networks are believed to be formed due to poor filling and packing of Cu around the percolated Cr-phase. These observations might be helpful to optimize the functional properties of Cu-Cr sintered alloys.

Highlights
- XCT coupled with machine learning-assisted segmentation
- Full 3D microstructural characterization of Cu-25Cr solid-state sintered composites
- Two porosity populations in Cu-25Cr composites: interfacial and matrix porosity
- Final density of Cu-25Cr composites is limited by interfacial porosity
- Cr-percolation, slow diffusion and thermal expansion mismatch inhibit strong interfaces

# 1. Introduction

Cu-Cr-based alloys show a desirable combination of mechanical, thermal, and electrical properties. Cu-Cr alloys with Cr content ranging from 5 to 50 %wt are mainly used as electrical contacts for vacuum interrupters for medium voltage applications. These properties are the result of a good electrical conductor, copper, and a semi-refractory metal, chromium, that can withstand the thermal constraints imposed by such applications [1-3]. Different manufacturing techniques have been used to produce these binaries Cu-Cr alloys, such as vacuum induction melting (VIM) [4], solid-state sintering [5-7], and more recently vacuum arc remelting (VAR) as well as powder bed fusion additive manufacturing techniques [8-10]. Because of the limited solubility of Cr in Cu, these alloys can be considered as Metal-Metal composites.

The microstructure of these alloys varies significantly from one manufacturing technique to another. On the one hand, the microstructures inherited from casting techniques result in a coarse cellular/dendritic primary Cr phase and a eutectic phase [4, 8]. On the other hand, sintering-based manufacturing techniques lead to microstructures that strongly depend on the Cr powder morphology and size and where the matrix can be considered as pure Cu [7, 23-24]. In addition to these previous microstructural characteristics concerning the Cr phase, the Cu/Cr interfaces are clearly one of the main differences [11-13]. It has been shown that for solidified Cu-Cr alloys, a semi-coherent Cu/Cr interface with misfit dislocations arrangements is established while for solid-state sintering alloys, the interface is incoherent and often characterized by the presence of vacancies between Cu/Cr, leaving pores at such interfaces [12]. This is a key aspect when it comes to the properties of composites-based materials. The physical and mechanical properties of composite materials are the result of the different matrix-reinforcement contributions, where interfacial nature and its properties make some of them. Therefore, the final properties of such materials may be partly disfavored by incoherent and weak interfaces [14-16].

The as-fabricated microstructure as well as the properties of these alloys have drawn the interest of several authors [1, 17-25]. It has been shown that these alloys' properties strongly depend on their composition and microstructure, especially regarding the size and morphology of the Cr-phase [17-25]. For Cu-Cr solid-state sintered alloys, an increased Cr content and the refinement of the Cr-phase will decrease the electrical and thermal conductivity but increase the hardness. Analytical models for predicting the functional properties of these alloys as a function of their microstructure are often used. The simple Rule-of-Mixture, the Hashin-Shtrikman model (Upper and Lower Bounds), Maxwell-derivative models, statistical-type models, Effective Medium Theory, and others [18-21] can be found in the literature. In addition, it has been shown experimentally and industrially that a Cr content from 25 to 35wt% shows the best compromise between properties and performances as electrical contacts for medium voltage applications, see [1, 5, 48]. The Cu-25wt%Cr is the most largely commercialized. However, most dedicated studies on the microstructure-properties relationships do not consider nor investigate in detail the porosity and Cr-phase distribution characteristics and their potential effects on the properties of Cu-Cr-based alloys. To this extent, X-ray Computed Tomography (XCT) has been proven a powerful non-destructive characterization method and has been extensively employed to investigate in three dimensions (3D) microstructural features, such as pore population and phase distribution [26, 27]. This is particularly helpful for metal matrix composites, which can suffer from polishing issues due to a softer matrix, and heterogenous distribution of second-phase particles which can be difficult to overcome by traditional 2D metallographic characterization. Therefore, microstructural features can be deeply investigated and statistically reliable within a representative microstructure volume. For example, key features such as the volume fraction, connectivity, and percolation of pores and phases can be analyzed. Thus, XCT brings new insights as a powerful tool for microstructural characterization.

In this study, the pore population, and the Cr-phase in Cu-25Cr solid-state sintered alloys have been characterized using XCT and deeply investigated using a specific image analysis routine developed in this work based on a machine learning segmentation approach. The pore population and Cr-phase distribution are characterized in samples showing different relative final densities. The size, morphology, spatial distribution, and connectivity of the pores as well as Cr-phase are reported and discussed for Cu-

25Cr solid-state sintered materials. Such pieces of information are crucial to understand the microstructure-property relationships of Cu-Cr composites. The methodology proposed in this work could also be employed to characterize other composite materials in the future.

## 2. Materials and Methods
### 2.1. Materials

Cu-25Cr alloy samples with three different final relative densities were fabricated by solid-state sintering under a secondary vacuum. Pure dendritic copper powder and pure irregular spheroidal chromium powder are mixed in a 3:1 mass ratio. The Cu powder has an average equivalent diameter of 40 µm and the Cr powder has a particle size distribution from 30 to 80 µm in equivalent diameter. To obtain three different final densities, three samples with different green-body densities were fabricated by varying the uniaxial compression conditions, named samples A, B, and C, respectively. The green densities of these samples were approximately A-85 ± 2%, B-89 ± 1% and C-91 ± 1%, respectively. Green densities were estimated geometrically by weighing green samples and measuring their dimensions using a digital micrometer. To maximize solid-state diffusion, the sintering heat treatment was done in a secondary vacuum near copper's melting point at 1050 °C as done in [7]. More details of the sintering parameters of Cu-25Cr materials can be found in reference [7]. All samples were sintered under the same conditions. Final densities were measured by Archimedes method in a Mettler Toledo Static Detect model XPR204 balance in triplicate for each sample. Figure 1a-b shows respectively the micrography of a typical microstructure of a Cu-25Cr solid-state sintered composite after a standard and specific polishing routine that reveals the porosity. The standard polishing procedure consists of mechanical polishing with 320 followed by 1200 SiC polishing paper. Surface finishing was done with 9, 3, and 1 µm diamond suspension, respectively, followed by a last step using a 0.1 µm colloidal silica suspension. The specific polishing procedure was identical to the standard one with the addition of a vibratory polishing with a 0.03µm colloidal silica solution for 10 hours. Optical micrographs were taken with an Olympus BX51RF optical microscope equipped with the Olympus Stream Essentials imaging software. Both the Cu-matrix and Cr-phase can be distinguished as well as some Cr-Cr inter-particle contacts. On the one hand, no pores can be seen in the microstructure in Figure 1a. As the Cu-matrix is very ductile, pores are filled with Cu during the mechanical process of surface polishing. This may lead to an erroneous interpretation and analysis of the resulting microstructure. On the other hand, in Figure 1b pores are revealed. Here, matrix pores and interfacial pores can be observed. This result brings a more correct interpretation and analysis of the microstructure of such materials. However, yet being a 2D characterization method, this experimental polishing procedure will hardly yield a representative observation of the whole porosity or Cr-phase fraction as well as spatial distribution in the volume.

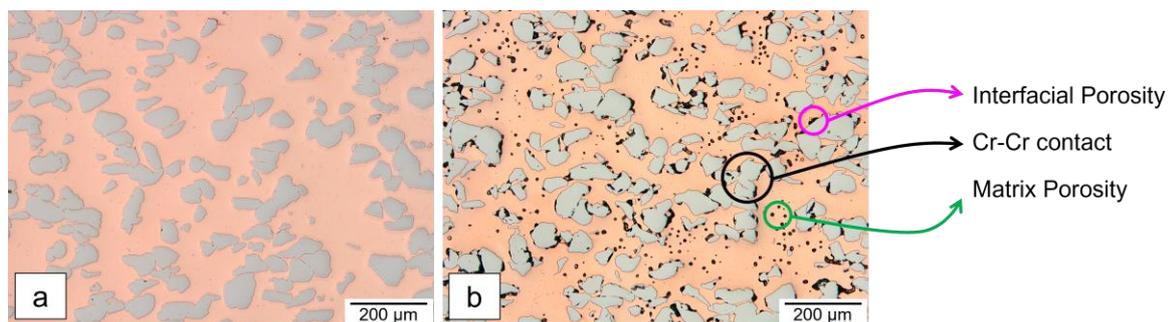

*Figure 1: Typical example of optical micrographs of Cu-25Cr solid-state sintered alloy after a) standard polishing procedure; b) specific polishing procedure revealing the porosity.*

## 2.2. X-ray Computed Tomography

### 2.2.1. Experimental Setup

X-ray computed tomography (XCT) was performed on cylindrical samples (7 mm in height and 1 mm in diameter). Samples were prepared by electrical discharge machining out of Cu-25Cr samples A, B, and C. An EasyTom XL tomograph from RX Solutions was used at 100 kV, with a 0.75 mm aluminum filter and 1.5 seconds acquisition per image with a total of 1792 projections. A volume of 1750×1125×1350 voxels was reconstructed by filtered back projection using XAct software from RX Solutions. Under these conditions, each experimental setup resulted in a voxel size of 0.93, 0.96, and 0.90 μm for samples A, B, and C, respectively. In this study, fine Cr precipitates and small pores of about 1 to 2 μm in diameter are considered the smallest microstructural features for the Cu-25Cr solid-state sintered materials. Therefore, the spatial resolution enables the detectability of pores as small as 3 to 4 μm in diameter and an accurate morphological quantification for pores as small as 5 μm in diameter.

### 2.2.2. Segmentation workflow

The same workflow for post-processing and analysis was applied to all the XCT data. The reconstructed images were cropped into a numerical cylinder of the same volume. Thus, all the samples had a corresponding cylinder of about 850 μm in diameter and 1200 μm in height for representing their microstructure. A 2-pixel radius 3D median filter was then applied to all the images followed by a 1-pixel radius 3D gaussian filter to reduce the noise. All these image analysis procedures were performed using Fiji [28]. Due to inhomogeneous grayscale level in the reconstructed image due to beam hardening effect as well as significant image texture and edge effects from the porosity to Cr-phase pixels, a simple threshold based on the gray level does not produce an accurate segmentation (Figure S1 in supplementary material). To overcome these segmentation issues, a machine learning-based routine was used to train a Random Forest pixel classifier with the consideration of image features, such as the grey pixel level, edges between different regions, and image texture. Image texture can usually be defined as a function of the spatial distribution of intensity or colors in a neighborhood of pixels to classify or define patterns inside an image.

The segmentation was performed with a pixel classification workflow available in ilastik open-source software [29]. Pixel classification workflow assigns labels to pixels based on pixel features and user annotations. For this work, pixel features such as smoothed pixel intensity, edge filters, and texture in the image were used with three labels. This allows the segmentation of all three phases with their respective pixel features, thus, the pores, the Cr-phase, and the Cu-matrix at the same time. Once the features are selected, a machine-learning random forest classifier is trained from user annotations interactively. It returns a probability map of each label from the interactively selected pixels. Then, each map is used to separate these labels into the desired microstructural features of Cu-25Cr material (Figure S2 in supplementary material). For more detailed information about the workflow and documentation of ilastik, see [29]. Due to the high pixel classification uncertainty at the interface between two phases, from a darker gray pixel's region (pores) to a medium gray region (Cr-phase) for example, an under and overestimation of the phase of interest was performed. This allowed a more representative pixel classification choice for final segmentation, especially for interfacial pores segmentation. Figures 2a-e and 2f-j show this process as an example of pores at the Cr/Cu interface and pores in the Cu matrix. Thanks to this machine learning routine not only the uncertainty and errors associated with the segmentation results can be estimated but a more significant representative segmentation of Cr-phase and pores can be performed. Post-segmentation image treatment consisted of applying a 1-median radius 3D pixel and 3-median radius 3D pixel filter in the segmented porosity and the Cr-phase, respectively, to eliminate single-noise pixels inherited from segmentation and smooth the outlines.

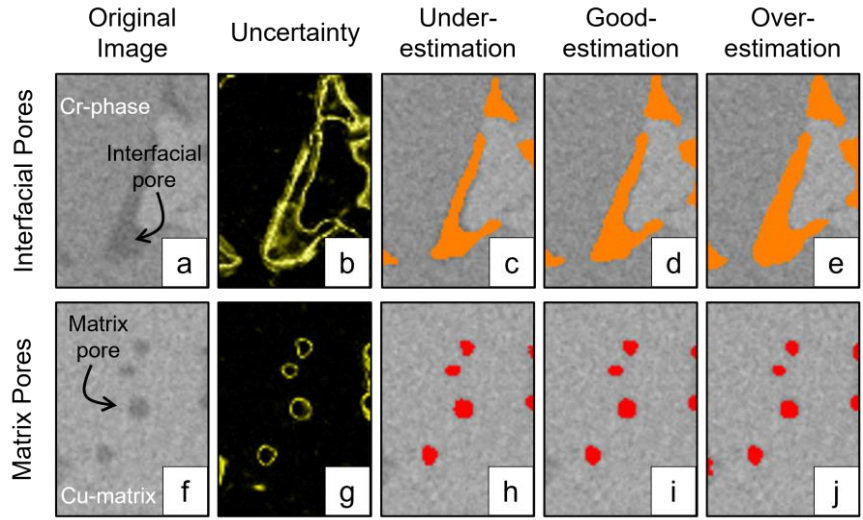

*Figure 2: a-b) and f-g) Examples of interfacial and matrix pores from the XCT image and its pixel classification uncertainty from the ilastik machine learning segmentation method. c-e) and h-j) Examples of an under, good, and overestimation for the porosity segmentation (orange for interfacial pores and red for matrix pores).*

### 2.2.3. Image analysis method and microstructural characterization

After the segmentation workflow, an image analysis of the segmented porosity and Cr-phase was performed in Fiji. For this, the plugin Analysis 3D [30] was used to label and characterize three-dimensional objects within the volume of interest. A 26-connectivity for a 3D voxel and a minimum volume of 8 voxels were used to identify and label each pore and Cr-phase.

A simple and conventional method based on sphericity and volume did not allow to properly discriminate both pore populations. This is nicely illustrated by Figure S3 provided in supplementary material. Thus, another methodology was required to improve the analysis. This was the motivation for developing the following algorithm. After labeling all the pores and all the Cr-phase particles in each sample, a specific algorithm to distinguish interfacial pores (those present in between Cr and Cu) from pores located in the Cu matrix was developed in Python, see Figure 3a-c. First, this algorithm identifies the surface of each identified object, see Figure 3b. Then, it creates a small 3D window with a pre-determined size that consists of a volume of interest analyzer (VOI analyzer). The latter will then follow only the surface of each labeled Cr-phase creating a search path, see Figure 3c. If any portion of a labeled pore is found inside this dynamic VOI analyzer, it automatically classifies this pore as an interfacial porosity. The size of the small window for the ROI analyzer was chosen by an optimization process to reduce computational time and increase precision in the selection process (see Figure S4a-b in supplementary material). A 3×3×3 voxel cubic VOI analyzer window was found to yield the best compromise and therefore it was used for each numerical sample. A 2D schematic representation of how the algorithm works is shown in Figure 3a-c.

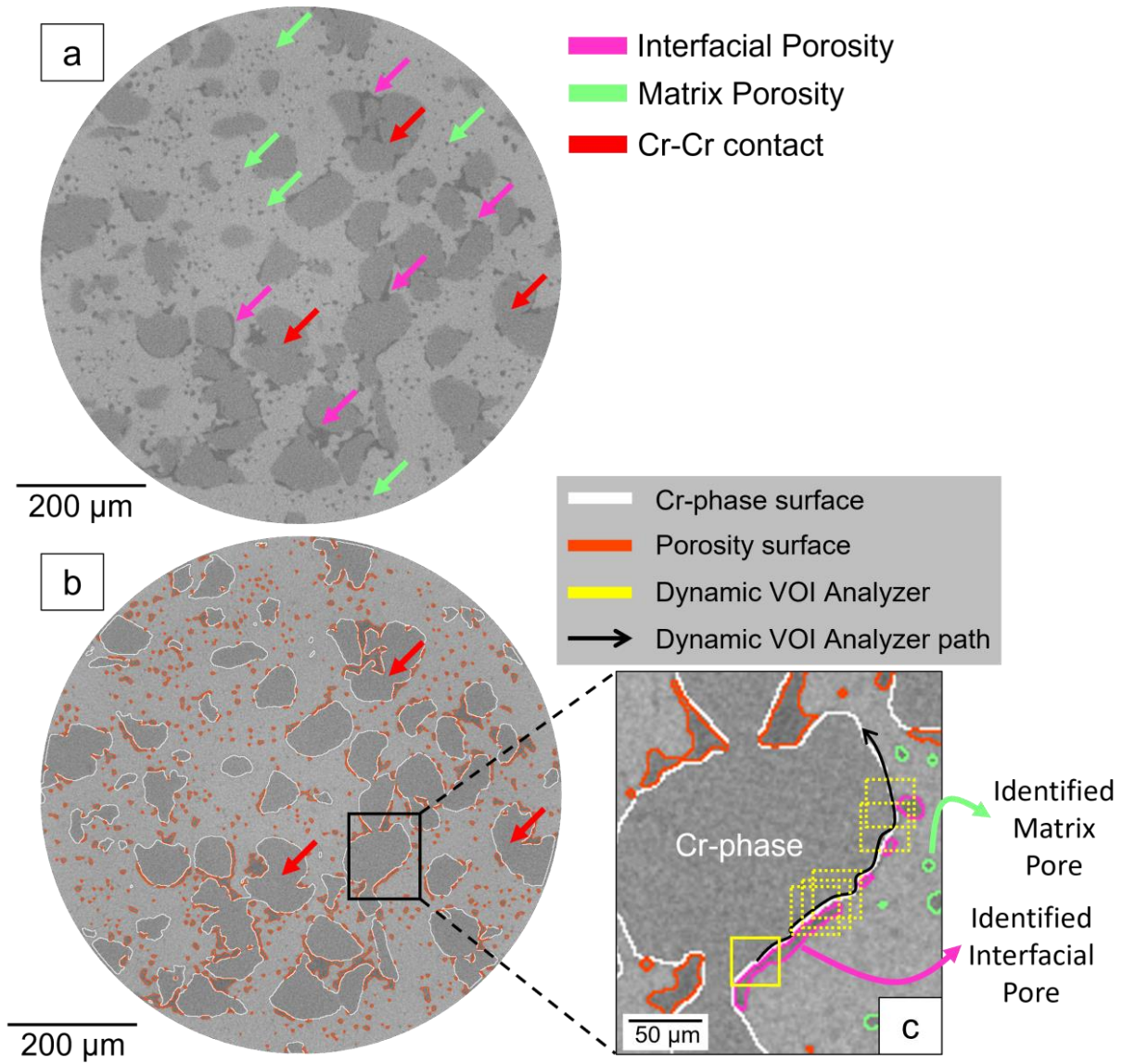

*Figure 3: a) 2D cross-section extracted from the 3D image showing the microstructure of sample A. Arrows indicate matrix (in green), interfacial pores (in magenta) and Cr-inter particle contacts (in red). b) The surface of each identified phase (pores in orange and Cr-phase in white) from the segmentation process overlayed with the XCT image in a). c)2D Representation of the dynamic VOI analyzer showing the search path on the Cr-phase surface and the pore's identification progress.*

Once all the Cr-phase surface of the segmented microstructure is run by this algorithm, interfacial pores and pores in the Cu matrix can be analyzed separately as well as the Cr-phase between sample A, B, and C. For porosity characterization, the volume, sphericity, and size of the bounding box in the local coordinates from the center of gravity (dx, dy, dz) of each pore are used in the analysis. This last is used to estimate the dimensions of pores in the directions perpendicular (x,y) and parallel (z) to the compression direction. From this, the aspect ratio of pores, dimension along the compression direction dz divided by the maximum dimension in the transverse direction (dx or dy). The maximum Feret diameter in 3D is used to evaluate the evolution of pore's size and their expansion. The angle between the Feret diameter axis of each interfacial pore and the compression axis is used to investigate their orientation with respect to the compression orientation. The volume of the largest pore divided by the total volume of pores for each sample is used as a relative volume to compare interfacial pores between the different samples. These parameters were obtained by the Analysis 3D and 3D Suite [31] Fiji plugin

For the Cr-phase characterization, the size and spatial distribution are analyzed. Due to the resolution limits of XCT image acquisition, the Cr-particles might be connected by the segmentation at various locations if Cr-inter particle contacts are present, as shown in Figures 1 and 3a-b. We have considered that the Cr-connections resulting from the segmentation are the result of effective Cr-inter particles contact, therefore, contributing to the geometrical percolation. Thus, the result of the 26-connectivity labeling without applying the disconnection algorithm is used to estimate the percolation of the Cr-phase. To disconnect and analyze the Cr-particles individually in the microstructure the algorithm 'disconnect particles' from Xlib Fiji plugin was used with a disconnection parameter k for 3D particles. This algorithm is a variation of the watershed algorithm (for more details see [32] and Figure S5 in the supplementary material). Then, labeling for the segmented Cr-phase after disconnection was made with the Analysis3D plugin. The size and sphericity as well as the average Cr to Cr distance from the first five neighbors for each labeled Cr particle are characterized to evaluate the spatial evolution from sample A to C.

## 3. Results

### 3.1. Validation of the image analysis workflow

A comparison between Figure 3a and Figure 3b shows that the machine-learning-assisted segmentation yields a significantly good identification of each microstructural feature. The interactive pixel classification combined with the evolutive random forest training enables a good distinction between Cr-phase and porosity. In addition, the under and over-estimation of the pixel classification allows a more representative choice for final segmentation. Therefore, the quantification of the overall porosity from this routine agrees well with the Archimedes method, see Table 1. This suggests that for all solid-state sintered Cu-25Cr samples studied in this work (relative density >93%), no open porosity is present. The Cr-phase separation due to inter-particle contacts and resolution limits is well handled by the 'disconnect particles' plugin (see an example in Figure S5). The algorithm developed to discriminate matrix pores from interfacial pores separates well these two pore populations. Figure 4a-b shows an example of the separated pores after the algorithm analysis for sample A. Interfacial pores, in magenta, are well separated from volume pores, represented in green. Figure 4c-d also shows a schematic illustration of the segmented microstructure of sample A after the application of the developed image analysis routine for a selected region of interest in the microstructure where the interfacial pores, Cu-matrix pores, Cr-phase as well as Cr-Cr inter-particle contacts are represented from a 2D to 3D perspective. The segmented microstructure in the x-y plane, which is a stack of single images in the z-direction (Figure 4c), can be represented in three dimensions by stacking all the images in the whole or selected volume of the numerical sample (Figure 4d). This can be plotted in a three-dimensional space where the 3D representation of the microstructure can be revealed, see Figure 4e.

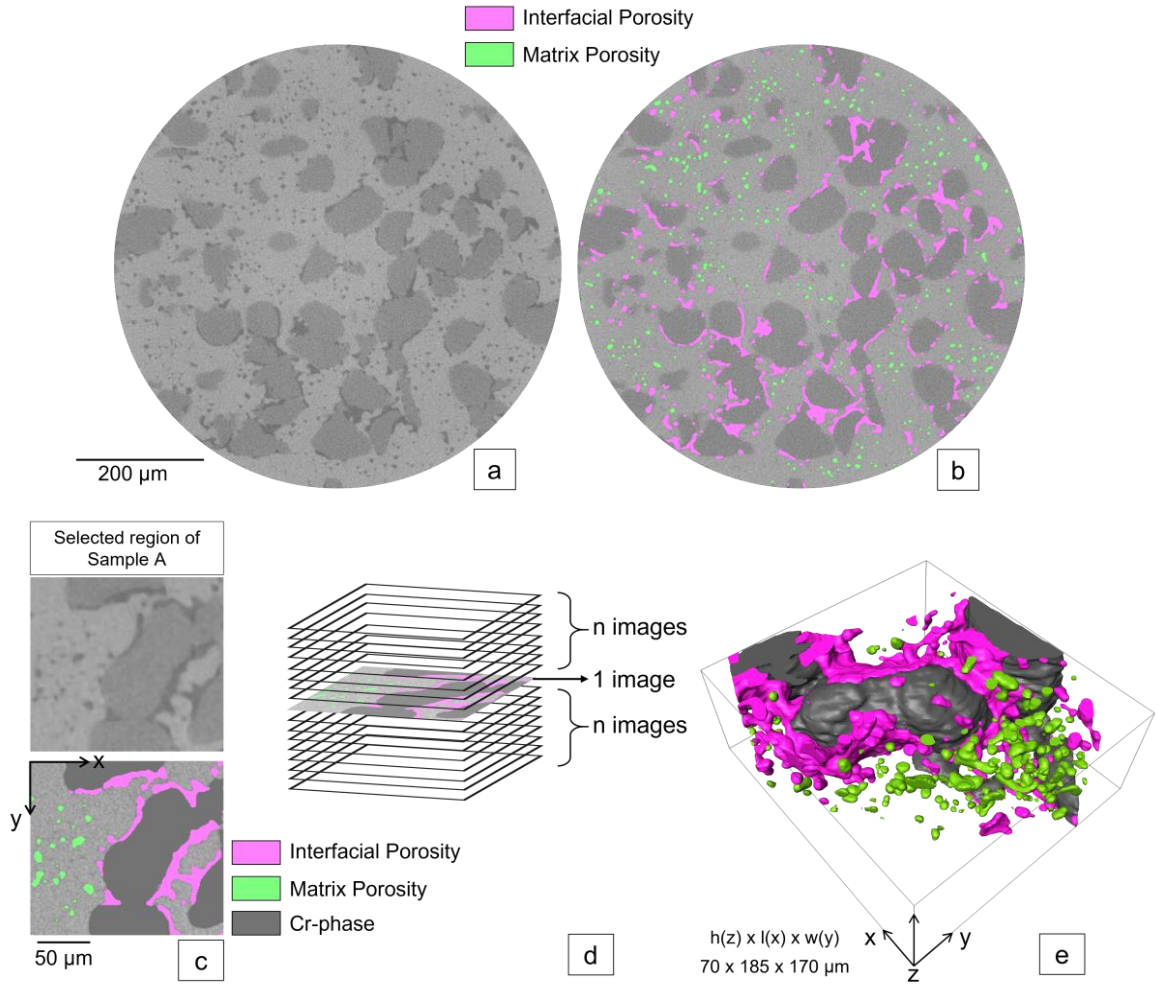

*Figure 4: a) 2D cross-section from the 3D reconstructed image showing the microstructure of sample A. b) Each identified and separated pore after the interfacial/matrix porosity separation algorithm overlayed with the XCT image in a). c) The selected region of interest in sample A shows matrix pores, interfacial pores, Cr-phase, Cr-Cr inter-particle contact, and its respective segmentation. d) Schematic representation of volume reconstruction after segmentation from several single cross-section images. e) 3D visualization of the Cr-phase, interfacial, and matrix pores. The bounding box size is shown in the lower left corner.*

### 3.2. Porosity quantification

The overall pore volume fraction as well as the interfacial and matrix pore's populations are characterized for samples A, B, and C after solid-state sintering (final density). The contribution of each pore population to the overall porosity is estimated. Figure 5a-c shows one cross-section image from the numerical cylindrical sample cropped after XCT reconstruction for each sintered Cu-25Cr material i.e., samples A-94% B-96%, and C-98%. From now, the final density of each sample determined using XCT is indicated along with the letter A, B, and C for the sake of clarity. Final relative densities for each sample from Archimedes method and XCT analysis are indicated at the top of each cross-section image. Table 1 summarizes the results of the calculated volume fraction of the segmented porosity by XCT analysis compared to the values determined using the Archimedes method. There is a good agreement between the relative density measured by the first method and the one calculated from the image analysis. Table 1 also gives the calculated interfacial and matrix porosity out of the total segmented porosity fraction, and their respective corresponding volume fraction. The higher uncertainty of sample A regarding the porosity fraction estimated based on the XCT reconstructed image is related to the larger pore population. As illustrated in Figures 2b and 2g, uncertainty is higher at the interface of the phase of interest from the classification method used by the machine learning workflow. Thus, when the interfacial and Cu-matrix pore fraction increases, the uncertainty of the overall pixel classification result also increases.

From the XCT images shown in Figure 5a-c, one can see that the porosity in sample A-94% is higher than in sample C-98%, by about 4 vol%. The differences in volume fraction of porosity between samples B-96% and C-98% are less obvious from the XCT 2D cross-section images when compared to the calculated values determined from XCT and Archimedes. The population of pores in the Cu-matrix decreases significantly as the relative density increases. This is clear when comparing samples A-94% with samples B-96% and C-98%. The density of interfacial pores and their size seem also to decrease as the relative density increases. Table 1 shows that there is an evident decrease in interfacial porosity from samples A -94% to C-98%. In sample C, a near fully dense Cu-matrix is observed, having about 0.1 vol% of pores located in the matrix.

Interparticle contacts between Cr-phase particles can also be seen in all samples. These contacts are present in all directions in the cross-section shown in Figure 5a-c as well as along the z-direction (parallel to the compression direction). The presence of interfacial pores in between very close Cr-particles even for highly compressed and densified samples such as sample C can be noticed, as indicated by the dashed black squares in Figure 5c. Overall, results show that the total volume fraction of porosity decreases in the sintered state with the increase of the initial compression pressure. Moreover, interfacial porosity represents the largest volume fraction of pores in all samples. As the final relative density increases, the corresponding fraction of interfacial porosity significantly increases, from about 74% in sample A-94% to 88% in sample B-96%, and finally reaching up to 95% of the overall porosity in sample C-98%.

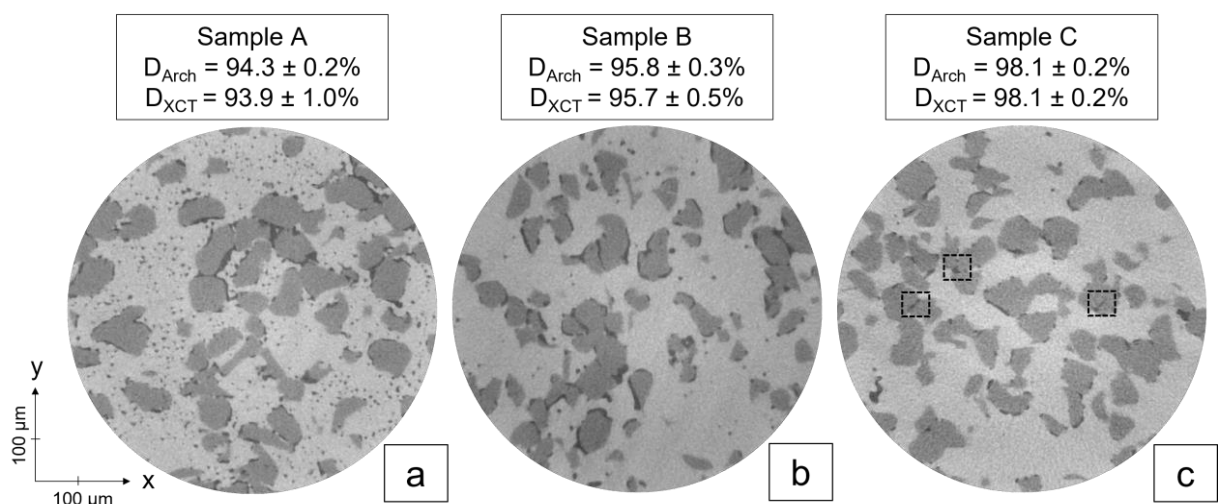

*Figure 5: 2D cross-section extracted from the 3D reconstruction for each Cu-25Cr solid-state sintered sample: a) sample A, b) sample B, and c) sample C. Pores are displayed in black, Cr-phase in dark gray, and Cu-matrix in light gray.*

*Table 1: Porosity volume fraction results from Archimedes and XCT, and after the algorithm for Cu-25Cr samples A, B, and C. The minimum and maximum density values from the triplicate measurements and the under and overestimation methods are represented by the symbol ±.*

| Cu-25Cr sample | Porosity Fraction - Archimedes (%) | Porosity Fraction - XCT (%) | Interfacial Porosity (%) | Corresponding Fraction | Cu-matrix Porosity (%) | Corresponding Fraction | Cr-phase Fraction – XCT (%) | Cr-phase Fraction – Theoretical (%) |
|---|---|---|---|---|---|---|---|---|
| **A** | 5.7 ± 0.2 | 6.1 ± 1.0 | 4.5 ± 0.6 | 74% | 1.6 ± 0.4 | 26% | 26.1 ± 2 | 27.7 |
| **B** | 4.2 ± 0.3 | 4.3 ± 0.5 | 3.8 ± 0.4 | 88% | 0.4 ± 0.1 | 12% | 24.7 ± 3 | 28.1 |
| **C** | 1.9 ± 0.2 | 1.9 ± 0.2 | 1.8 ± 0.15 | 95% | 0.1 ± 0.05 | 5% | 30.1 ± 2 | 28.9 |

*3.3. 3D characterization of matrix porosity*

The morphological and size characteristics of the pores located in the Cu-matrix for samples A-94%, B-96%, and C-98% are shown in the sphericity-volume dispersion plot displayed in Figure 6, respectively. As final density increases due to a higher green density, matrix pores tend to show higher values of sphericity and slightly smaller volumes, generally above 0.7 and under $10^3$ µm³. This is clear for samples B-96% and C-98% where the Cu-matrix is nearly dense. This tendency is evident by the shift in sphericity to higher values and volume to smaller values as shown in Figure 6. On the one hand, sample A-94% presents a large population of pores which consists of small and spherical pores as well as significantly large pores having a complex branch-like morphology. On the other hand, as the green density increases and consequently the final relative density also increases, these large pores evolve into smaller and less complex pores, gaining in sphericity. Thus, sample C-98% presents a smaller but more spherical population of pores. The morphology and size evolution from sample A-94% to C-98% are illustrated using the 3D representation of a matrix pore located at the end of each population in the dispersion plots.

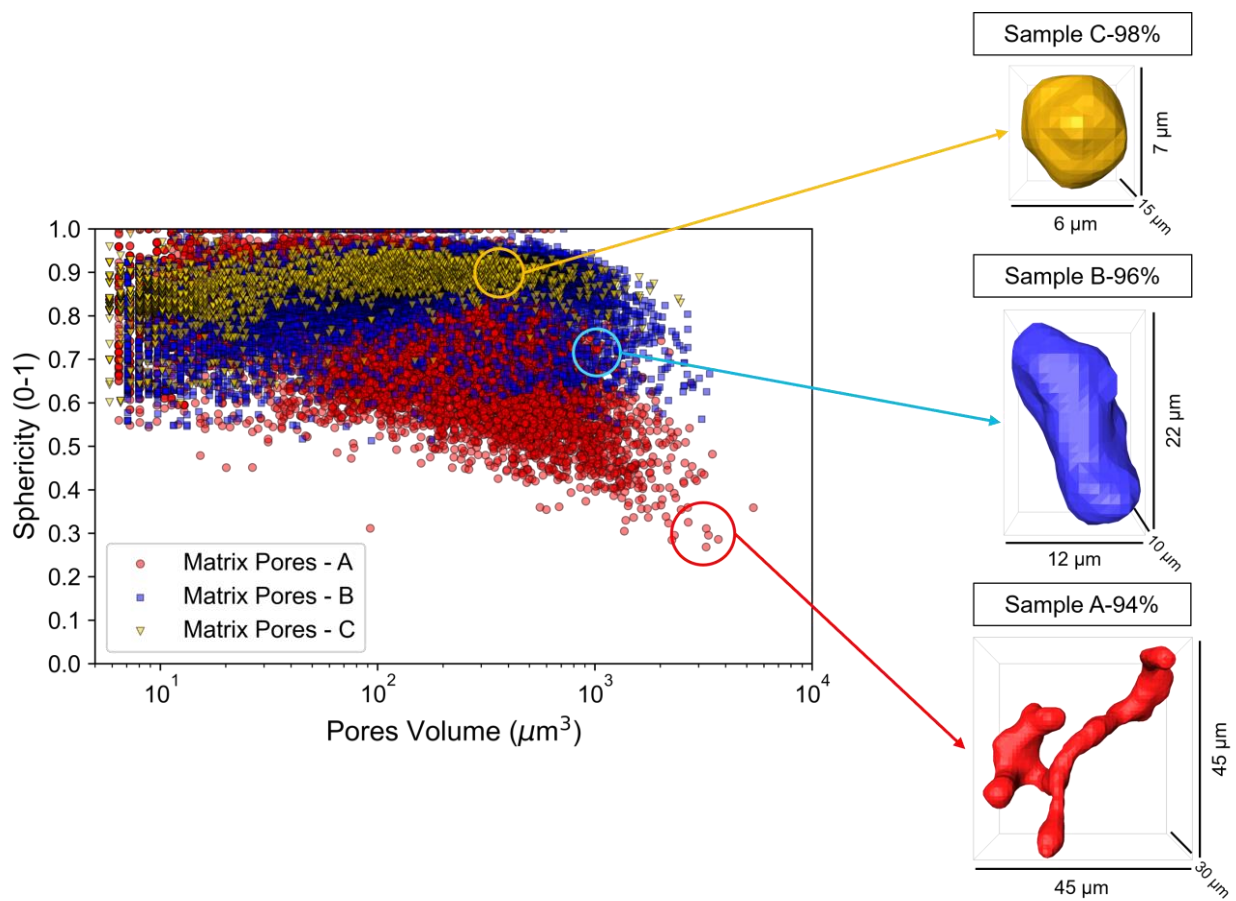

*Figure 6: Porosity Sphericity - Volume dispersion plot for matrix pores for the different Cu-25Cr samples, sample A-94%, sample B-96%, and sample C-98% and one representative 3D plot from each pore population from the region demarked in the dispersion plot.*

*3.4. 3D characterization of interfacial porosity*

Interfacial pores are characterized by their aspect ratio, volume, and maximum Feret diameter. Differently from matrix pores, interfacial pores can reach large volumes in the microstructures, up to approximately 3.2% of the total volume of interfacial porosity. In addition, interfacial pores can reach large dimensions within Cu-25Cr solid-state sintered materials. Figure 7a-b shows respectively the aspect ratio of interfacial pores between the dimension along the compression direction and the transverse dimension (perpendicular to the compression axis) as well as the maximum Feret diameter of each interfacial pore. This allows the comparison of the size of each pore in the volume of the sintered materials as well as in which direction there are mostly expanded. The largest interfacial pores reach larger dimensions in the transverse direction, i.e. perpendicular to the compression direction (aspect ratio less than 1) as shown in Figure 7a. No difference from sample A-94% to C-98% is evident regarding the aspect ratio. From Figure 7b one can see by the large Feret diameters that interfacial pores can expand and connect within the microstructure reaching a significant volume fraction of the overall volume fraction of interfacial porosity. Pores over 1 vol% of the overall volume fraction of interfacial porosity can expand up to 550 µm in size into the microstructure. Therefore, large and expanded networks of interfacial pores can be found within the microstructure, especially in samples A-94% and B-96% In addition, interfacial pores are present in the microstructure with a certain flatness orientation with respect to the compression direction regardless of the relative density. For all samples, there is a significant concentration of pores with an aspect ratio lower than 1 as Figure 7a shows. Moreover, the angle between the maximum Feret diameter (the highest dimension of each interfacial pore) and the compression axis shows a cumulative distribution near high angles (over 60°), confirming this trend, see Figure 7c. Especially for the largest interfacial porosity, this flatness tendency is more evident.

As the final relative density increases, there is a decrease in Feret diameter and relative volume. This suggests that interfacial pores are smaller and less connected and expanded. This tendency is clear from samples A-94% to C-98% where the largest pore decreases significantly in volume and in Feret diameter, from 3.2 vol% and 550 µm to 0.7 vol% and 200 µm, respectively. From these results, one can say that from samples A to C, higher green and final density leads to the formation of smaller and less expanded interfacial pores. The dashed lines in Figure 7b indicate the range of this effect from sample A-94% to B-96% and B-96% to C-98% respectively.

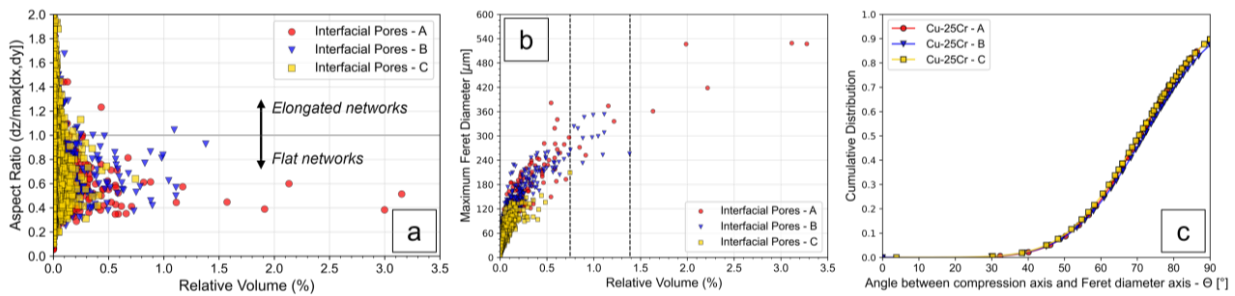

*Figure 7: a) Aspect ratio (ratio between the size of the bounding box along the compression direction and the maximum dimension along the transverse direction) for each interfacial pore as a function of their relative volume for Cu-25Cr samples A-94%, B-96%, and C-98%. b) Maximum Feret diameter for each interfacial pore as a function of their relative volume for Cu-25Cr samples A-94%, B-96%, and C-98%. These values serve as interfacial pore expansion within the microstructure .c) Cumulative distribution of the angle between the compression axis and the Feret diameter axis for each interfacial pore.*

The number of interfacial pores in this range increases whereas the corresponding volume fraction out of total interfacial porosity remains approximately the same. For sample A-94%, the 5 interfacial pores shown in Figure 8a correspond to around 12 vol% of the total interfacial porosity. As for sample B-96%, the 13 interfacial pores displayed in Figure 8b correspond to about 13 vol%. For sample C-98%, 52 interfacial pores are required to reach 12 vol% of the total interfacial porosity. Figure 8a-c shows the interfacial porosity fraction mentioned previously in the whole volume of the numerical sample for the 3 conditions investigated. Each interfacial pore is represented by one color. The large network of interfacial porosity and the mentioned effect are then clearly revealed in the three samples. The large branched interfacial pores networks and the flatness orientation previously mentioned are evidenced in Figure 8. Furthermore, the volume of the first 20 interfacial pores is shown in Figure 9, from the largest to the smallest. The effect of higher green and final density on the formation of these interfacial pores is clear.

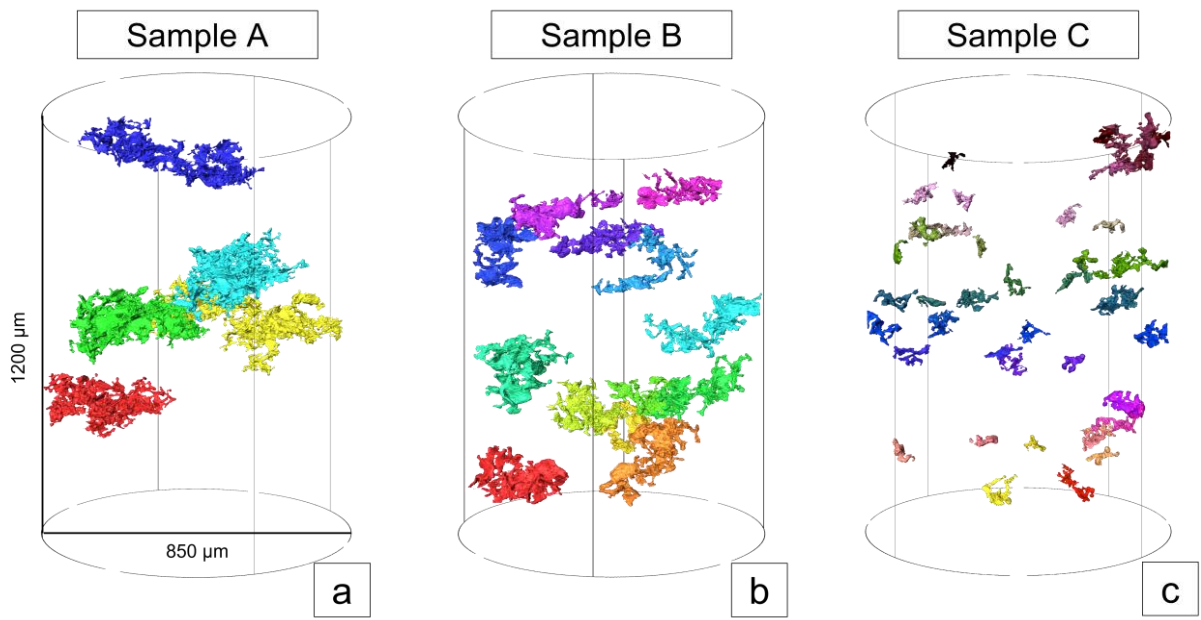

*Figure 8: Volumetric representation of the largest interfacial pores up to the range indicated by the dashed lines in Figure 7b, a) 5 large interfacial pores represent 12vol% of the total interfacial porosity for sample A , b) 13 large interfacial pores represent approximately 12vol% of the total interfacial porosity for sample B,, and c) 52 interfacial pores in sample C correspond to 12vol% of the total interfacial porosity. Each pore is represented using one single color inside the cylindrical volume.*

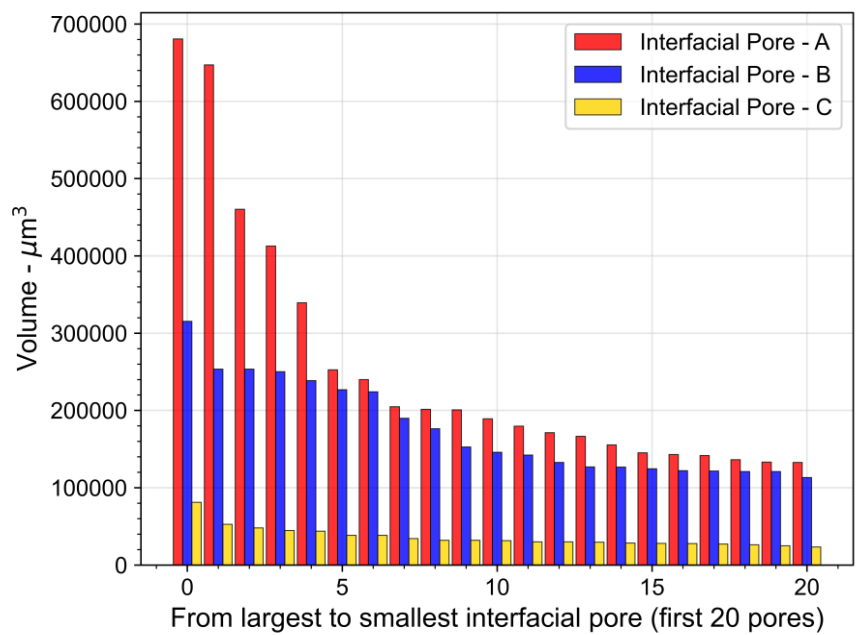

*Figure 9: Volume of the first 20 Interfacial Porosity, from the largest to the smallest.*

Figure 10a-c shows the largest interfacial pore in-between the percolated Cr-phase (in transparent gray) for samples A-94%, B-96%, and C-98%, respectively. The presence of each pore in the surroundings of Cr particles is rather obvious. In Figure 10a, regions of the large interfacial porosity form a cup shape, resulting in a pore envelope around Cr particles. Here, interfacial pores can represent a significant volume fraction locally in the microstructure, interconnecting between Cr-percolated particles.

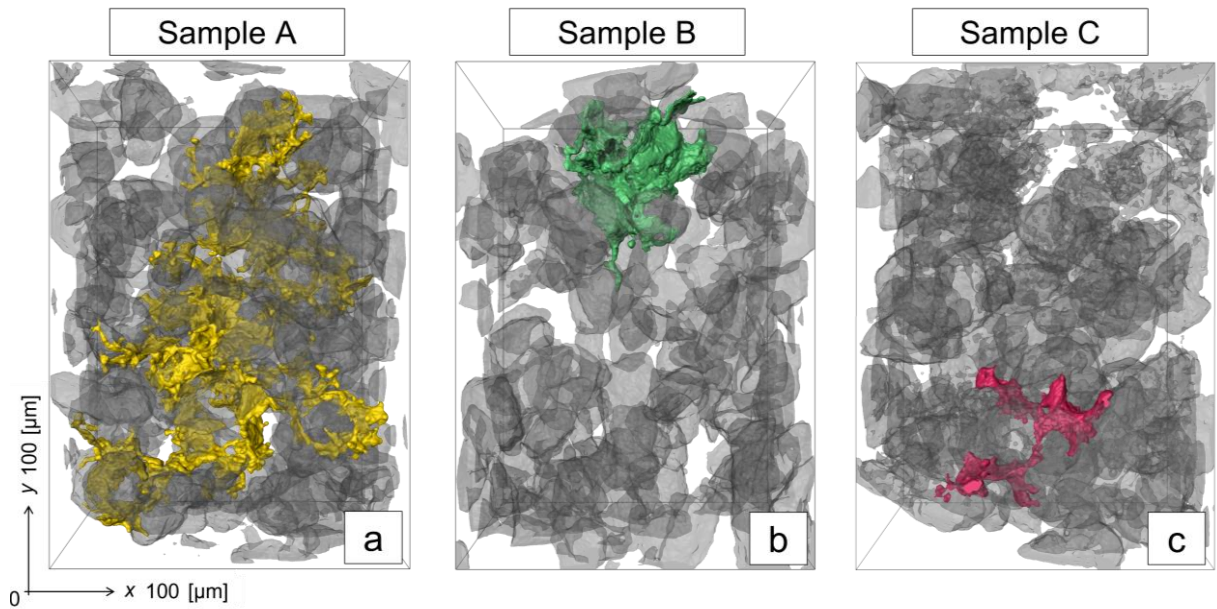

*Figure 10: 3D visualization of the largest interfacial pore in-between the percolated Cr-phase for Cu-25Cr sample a) A in magenta color, b) B in orange color, and c) C in red color. The percolated Cr-phase around the pores is in light-transparent gray. Here in Figure 9a-c, the color of each pore is the corresponding color of Figure 8a-c.*

*3.5. 3D characterization of the Cr-phase*

The Cr-phase volume fraction estimated based on the XCT image segmentation is also given in Table 1 along with the theoretical value. Here, the theoretical value is considered to be the volume fraction of 25 wt.%Cr in a Cu matrix for a material having the density found using the Archimedes method. The equation used to calculate this theoretical value can be found in the supplementary materials. Therefore, if Cu-25Cr materials were fully dense, the Cr volume fraction would be around 29.4 vol%. For each sample, as porosity is present in all the samples, the Cr volume fraction decreases slightly as porosity increases. Values found based on XCT images agree well with the theoretical ones. Some differences can be found, especially for samples B-96 and C-98%, due to some variation in gray level intensity from pixels in the center and border of the sample. Therefore, the pixel classification may not be as precise as for porosity.

Morphological characterization of the disconnected Cr-particles is shown in Figure 11a. The sphericity versus Cr-particle equivalent diameter, calculated from the volume of each particle, does not show significant differences for the three materials. Cr-particles segmented from the XCT and image analysis routine and separated by the 'disconnect particles' algorithm present an equivalent spherical diameter and sphericity in the range of the original powder. For sample C-98%, a tail-like shape dispersion can relatively be seen for larger equivalent diameters, i.e. larger volumes. In addition, Figure 11b shows the average Cr to Cr particle distance for the first five neighbors of a Cr particle in the microstructure. As there is an increase of 6% in green body density from samples A to C (green densities: A-85%, B-89%, C-91%) that produces different final densities (final densities: A-94%, B-96%, C-98%), one can suggest that the 3D distribution of Cr particles can be altered. However, there are no significant differences between the three Cu-25Cr solid-state sintered composites produced by different compaction pressure. Most of the Cr-particles distance one another by 60 μm. Only a slight shift to smaller average distances is notable for sample C-98% (the one exhibiting the highest relative density).

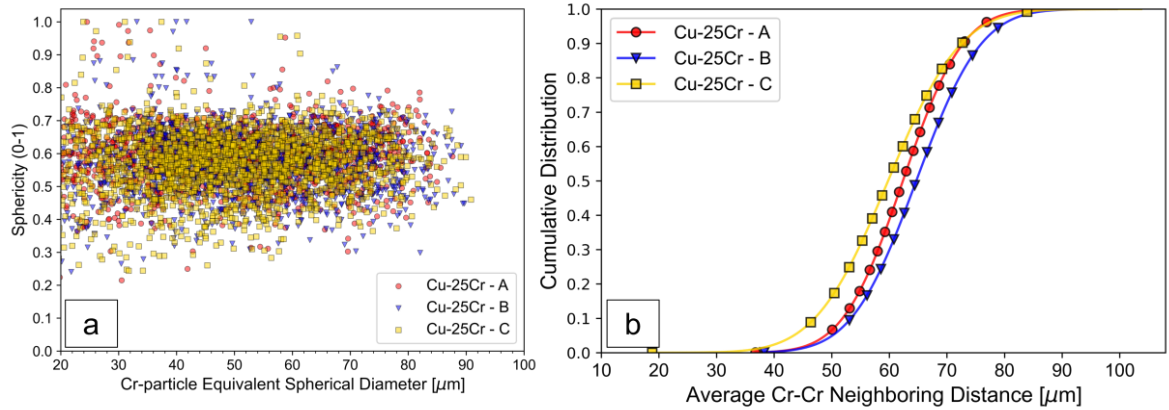

*Figure 11: a) Porosity Sphericity - Volume dispersion plot for Cr-phase after disconnecting Cr-Cr particles in contact for Cu-25Cr samples A-94%, B-96%, and C-98%. b) Cumulative distribution of the average Cr to Cr distance for the five closest neighbors of a Cr particle.*

Figure 12a-c shows the 3D volume representation of the estimated percolation of the Cr-phase by the labeling method previously mentioned without disconnecting the Cr-particles. A large volume of single-color Cr particles can be identified. Therefore, for the large volume fraction of blue-colored Cr particles, inter-Cr-Cr contacts are present, resulting in Cr-phase geometrical percolation. Therefore, this suggests that the Cr-phase in Cu-25Cr solid-state sintered materials presents a significant volume fraction of the Cr-phase in a percolated configuration, i.e. Cr inter-particle contact coordination. One can estimate the volume fraction of the larger Cr percolated volume: 82% in sample A-94%, 85% in sample B-96%, and 91% in sample C-98%. Thus, there is a relative increase in the volume fraction of the percolated Cr-phase from samples A to C. As green and final density increases, there is a slight increase in the Cr-percolated volume with no significant spatial change across Cr-Cr neighbors.

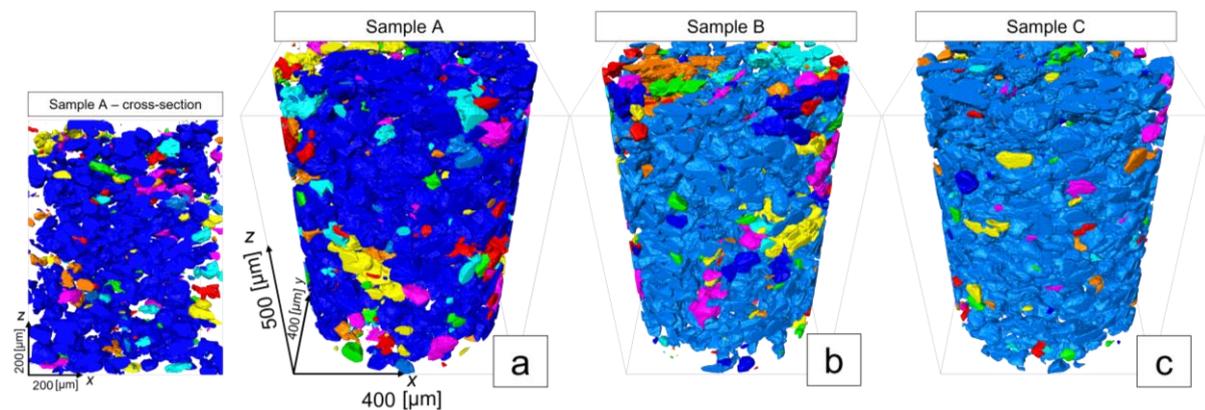

*Figure 12: 3D visualization of the Cr-phase percolation by labeling the Cr-phase with a 26-connectivity for samples a) A-94% with an additional cross-section visualization; b) sample B-96%, and c) sample C-98%. Each color represents one single 'particle'. Therefore, Cr particles with the same color can be considered in contact and therefore percolated:*

## 4. Discussion

### 4.1. On the matrix porosity evolution

The driving force for solid-state sintering is the decrease of surface free energy of compacted powder. For metal matrix composites, the densification process and the sintering mechanisms strongly depend on the sintering mechanisms of the matrix, in this case, copper. In general, for spherical and dendritic copper powders, at high temperatures ($T_{sintering} > 0.9T_{Melting}$), the predominant mechanism for densification is grain boundary diffusion and volume diffusion from boundaries [33, 34]. For Cu-Cr-based materials, these mechanisms are also considered the mechanisms responsible for densification [7,36]. Therefore, the initial green body density (the density after compression prior to sintering), especially in the matrix, is an important variable to consider when optimizing a sintering process and the final density of sintered composite materials made of soft and hard metallic phases. To some extent, the increase of the green body density increases the final density. This is because with the increase of green body density, matrix pores are already reduced and potentially closed before densification [36]. This is clear for samples A, B, and C studied in this work. Figure 5a-c and Table 1 clearly show that with the increase of the compression pressure, the green body density gets improved, and this goes along with an increase in final relative density. This is clear for the reduction in porosity within the Cu-matrix from samples A-94% to C-98%. From the dispersion plots in Figure 6a-c, there is a reduction in number as well as in volume for Cu-matrix pores. Matrix pores become smaller and more spherical from samples A-94% to C-98%, which agrees well with the evolution of densification of the Cu-matrix. Therefore, the higher green body density results in a higher final relative density for the Cu matrix for the applied sintering conditions. Porosity before sintering is closed and smaller due to higher compression and compaction resulting in a nearly fully dense Cu-matrix in sample C-98% with a smaller and spherical Cu-matrix porosity.

*4.2. On the interfacial porosity evolution*

The green body density for solid-state sintered composite materials in which the second phase is less ductile than the matrix and non-reactive can play an important role in the matrix-inclusion interaction during compression and densification [37-42]. Adding 25wt.%Cr results in a sufficient volume fraction to overcome the geometric percolation threshold of the Cr phase in the Cu-Cr metal-metal composite [35]. The Cr inter-particle contacts are evident in Figure 5a-c. In this case, at the temperature employed to densify the Cu-Cr materials, Cr-Cr sintering does not play a role [36]. It has been shown that hard inclusions in a ductile matrix result in high plastic strain in the Matrix/Inclusion interface and its vicinity [37-41]. The morphology of the inclusion will also influence the strain state as well as the packing and filling of the matrix powder in the inclusion neighborhood [41]. This can lead to a higher density of pores at the matrix/inclusion interface. This is the case for Cu-Cr-based solid-state sintered composites. Cr-powder has an irregular morphology and is less ductile than the Cu-matrix. Therefore, during compression and compaction, the dendritic Cu-powder does not pack efficiently on the surface of the Cr-inclusions, especially at the extremities of Cr-particles perpendicular to the compression direction. In addition, the presence of local Cr dispersion heterogeneity as well as percolation of the Cr-phase can play an important role in the filling and packing of Cu powder in the percolated inter-Cr spacing [35, 42]. The Cr-phase percolation was identified by the 26-connectivity labeling mentioned previously. We consider that these Cr-connections resulting from the segmentation are the result of an effective Cr-inter particles contact, therefore, contributing to the geometrical percolation of Cr. Therefore, the percolation volume was estimated before applying the disconnect particles. Figure 12a-c shows only the segmented Cr-phase for each sample in 3D visualization. The inter-Cr-Cr contacts are clearly present, resulting in Cr-phase geometrical percolation. Thus, a significant volume fraction of the Cr-phase percolates in the microstructure even for a low green and final density Cu-25Cr material. As green and final density increases, there is a slight increase in the Cr-percolated volume with no significant spatial change across Cr-Cr neighbors as mentioned previously. Therefore, this increase may be only due to a small Cr-Cr rearrangement and some development in Cr-Cr clustering rather than significant spatial changes for Cr particles.

Figures 8a-c and 9 also show that there is a decrease in the largest porosity size and volume from samples A-94% to C-98%, from 3.2% to 0.7% as for the relative volume of total interfacial porosity. This is clear from Figure 7b where the relative volume and the maximum Feret diameter of interfacial pores from samples A-94% to C-98% reflect the formation of smaller and less expanded interfacial pores by the increase of green and final density. This can be seen with the aid of the dashed lines in Figure 7b. While sample A-94% presents a very large and well-connected interfacial porosity, sample B-96% and sample C-98% present a significant decrease in volume and expansion. From samples A-94% to B-96% and from B-96% to C-98%, the largest relative volume is decreased by a factor of two and the Cr-percolated also increases. Since the Cr-phase is not significantly affected morphologically and spatially by the initial difference in green density, nor by the sintering process, the main responsible mechanism for this trend is thought to be the better filling and packing of Cu around Cr-particles.

We suggest that the high density of Cr inter-particle contacts and the percolation of the Cr-phase contributes to forming voids at the Cu/Cr interfaces and its complex Cr-clustering vicinity during compression. This means that the lack of filling in the vicinity of Cu-Cr interfaces, especially in the transverse direction of compression leads to the formation of these interfacial pores. Those interfacial pores expand and interconnect in a plane perpendicular to the compression direction (relatively flat morphology) as shown in Figures 7-10. Moreover, the diffusion of Cr in Cu at high temperatures is of the order of magnitude of $6.10^{-10}$-$5.10^{-9}$ cm²/s [36, 43-45]. Thus, the typical diffusion lengths are relatively low, around 30 to 50 µm at the sintering temperature. This does not contribute to the elimination of interfacial pores during sintering nor to produce strong bonded interfaces. It is also worth mentioning that thermal expansion may also play a role in the generation of interfacial pores. Since the coefficient of thermal expansion of Cu and Cr are significantly different, copper's is about three times higher than chromium's [46,47], high mechanical stresses can be generated between Cu and Cr during the heating and cooling steps of the thermal sintering process. Consequently, the interfacial porosity between Cu/Cr strongly survives sintering and densification. The increase of compression and compaction state seems, thus, to better fill and pack the inter-Cr spacing of the percolated Cr-phase with the Cu-matrix as well as develop more the Cr-clusters (i.e., more Cr-Cr contacts). This result in a decrease of size and connection of the largest interfacial pores as well as to significantly increase Cu-matrix density.

*4.3. Possible consequences of large interfacial pores and Cr-percolation on the microstructures-property relationships*

These results suggest that the final relative density of Cu-25Cr solid-state sintered materials is limited by interfacial porosity. A highly dense Cu-matrix is reached with an overall relative density of 96% leaving the interfacial porosity as the sole representative volume fraction of porosity. The increase of compression of the green body seems to enable a gain in relative density. However interfacial porosity, especially in the region where the Cr-phase is percolated, is still present. This interfacial porosity network can play an important role as a third phase. Samples A and B show some large interconnected interfacial pores that can expand within the microstructure and, thus, have a significant volume fraction locally. In addition, Cr-percolation and 3D distribution appear also to be important aspects to understand the properties of Cu-Cr composites as well as local sintering behavior. The spatial distribution and percolation of particles in particulate composites are known to strongly influence the properties [49-51]. For hard particles embedded in a soft matrix, the percolation of the hard phase can lead to limited densification of the green body during compression loading due to the formation of a hard and resistant 'skeleton'. Thus, the compression and compaction of green bodies will be mostly limited by the deformation of the softer matrix until reaching the maximum load. For physical properties which depend on conduction, such as electrical and thermal conductivity, the presence of a large and connected network of non-conductive phases as well as the percolation of a second phase is critical for the final properties of the composite material [14-17]. Conduction path and scattering events will change for carriers, such as electrons, in the presence of a percolated and non-conducting phase. The mechanical properties of such composites will also be affected by the presence of these large interfacial pores and second-phase percolation due to a limited load transfer from the soft matrix to the hard reinforcements.

This motivates and encourages further studies to establish more detailed microstructure-properties relationships of Cu-Cr-based solid-state sintered materials, especially considering the porosity fraction as well as the nature and properties of the interfaces.

## 5. Summary and Conclusion

In this study, we investigated the pore population of Cu-25Cr solid-state sintered alloys with three different relative final densities, A-94.3%, B-95.8%, and C-98.1%. XCT coupled with a machine-learning-assisted segmentation has proven to be an efficient and precise characterization method that allows a full 3D characterization of different microstructural features overcoming the issues associated with more simple segmentation strategies. The performed machine learning-assisted segmentation as well as the dedicated image analysis routine allowed the discrimination and quantification of matrix pores from interfacial pores. The results of the porosity volume fraction estimated based on this workflow agree well with the results obtained by the Archimedes method. The Cr-phase volume fraction obtained by XCT also agrees well with theoretical values. For all samples, interfacial porosity represents undoubtedly the larger fraction, from 74% of the total volume of pores in sample A and up to 95% in sample C. Matrix pores tend to be closed more easily with the increase of compression of the green body, resulting in a near fully dense Cu-matrix for relative final densities above 96%. The presence of Cr-percolation combined with its irregular morphology, its diffusion kinetics, as well as its difference in coefficient of thermal expansion in comparison with the Cu matrix, does not promote the formation of a strong interface. It is also believed that it does not favor the filling and packing of Cu-powder during compression and compaction around the percolated Cr-phase, especially in the direction perpendicular to the compression axis. Thus, this suggests being the main cause of large and connected networks of interfacial porosity. The increase of the compaction state of the green body can lead to the formation of smaller and less expanded large interfacial porosity networks, reducing its size and volume by the better filling of Cu around Cr-particles. Thanks to this dedicated routine the 3D characterization of the microstructure of sintered Cu-25Cr composites the large volume fraction of interfacial porosity will certainly be helpful to shed light on the differences in properties between sintered and cast products that do not contain as much porosity and where interfaces are stronger. The developed image analysis routine may also be applied to other cases where 3D microstructure characterization is required.

**Data availability**
Data is still being used as part of another study. Reasonable requests for the data can be made to the corresponding author in the meantime.

**Acknowledgments**
The authors are grateful to Schneider Electric Industries and Association Nationale de la Recherche Technologique (ANRT) for the financial support of this work in the framework of the PhD of Lucas Varoto (CIFRE 2021/1619). The authors are also grateful to Stephane Coindeau, from the Grenoble INP - CMTC characterization platform, for the technical and experimental support of the XCT experiments.

**Supplementary Materials**
Details about the advantages of XCT applied to Cu-Cr solid-state sintered alloys, pixel classification routine and other supplementary results can be found in the attached document entitled "Supplementary Materials".